\documentclass[12pt,leqno]{amsart}

\usepackage{amsmath,amssymb}

\def\Ee{{\mathrm{e}}}

\def\EE{{\mathbb{E}}}
\def\CC{{\mathbb C}}
\def\RR{{\mathbb R}}

\def\SO{\mathrm{SO}}
\def \dalpha{{\dot \alpha}}
\def \dbeta{{\dot \beta}}

\def \Cliff{{\mathrm{Cliff}}}
\def \CLIFF{{\rm{CLIFF}}}

\def \SPIN{{\mathrm{SPIN}}}

\def \bee{{\mathbf e}}

\def \Ker{{\mathrm{Ker}}}

\def \ii{{\sqrt{-1}}}
\def\Not{\not \!\!}

\def\pt{{pt}}
\def\overwave{\tilde}

\def \sa{{\mathrm{sa}}}

\newtheorem{definition}{Definition}[section]
\newtheorem{theorem}{Theorem}[section]
\newtheorem{proposition}{Proposition}[section]
\newtheorem{corollary}{Corollary}[section]
\newtheorem{remark}{Remark}[section]
\newtheorem{lemma}{Lemma}[section]

\def\dfrac#1#2{{\displaystyle\frac{#1}{#2}}}

\def\book#1{{\em{#1}}}
\def\paper#1{{\em{#1}}}
\def\jour#1{\textrm{#1},}
\def\yr#1{({\textrm{#1})}}
\def\vol#1{\textrm{#1}}
\def\pages#1{\textrm{#1}}

\title{Submanifold Dirac Operator with Torsion}
\author{Shigeki Matsutani}
\date{}
\begin{document}

\maketitle

\begin{abstract}
The submanifold Dirac operator has been studied
for this decade, which is closely related to
Frenet-Serret and generalized Weierstrass relations.
In this article, we will give
a submanifold Dirac operator defined
over a surface immersed in $\EE^4$
with U(1)-gauge field as
torsion in the sense of the Frenet-Serret
relation, which also has data of immersion of
the surface in $\EE^4$.
\end{abstract}



\section{Introduction}

The submanifold quantum mechanics  was opened by
Jensen and Koppe \cite{JK} and de Casta \cite{dC}.
The submanifold Dirac equations were
studied in \cite{BJ, M1, M2, M3, M4, M5, M8, MT},
which are closely related to recent movements in
differential geometry. The same
Dirac operator as one in \cite{M5}
appeared in a lecture by Pinkall in ICM of 1998
\cite{PP}, in which  conformal surfaces in the
euclidean space were studied using the Dirac operator.
The related Dirac equation is known
as the generalized Weierstrass relation in the category
of differential geometry \cite{Ko, PP, Ta}.

Recently, this author gave an algebraic construction of the
submanifold quantum mechanics,
which  exhibits nature
of submanifold \cite{M7, M8}.

On the other hand,
 for a space curve in three dimensional euclidean
space $\EE^3$, Takagi and Tanzawa found
a submanifold Schr{\"o}dinger operator with a gauge field
\cite{TT},
\begin{equation}
	{\mathcal S} :=-(\partial_s-\sqrt{-1}a)^2
- \frac{1}{4}k^2, \label{eq:S-1}
\end{equation}
whereas the original one in \cite{dC, JK} is given by
\begin{equation*}
	{\mathcal S} :=-\partial_s^2
- \frac{1}{4}k^2,
\end{equation*}
where $k$ is a curvature of the curve.
The existence of the gauge field $a$ is due to the fact
that the codimension is two.

In this article, we will generalize the Takagi-Tanzawa
Schr{\"o}dinger operator (\ref{eq:S-1}) to the submanifold
Dirac operator over a surface
in $\EE^4$ using the algebraic construction.

After giving a geometrical setting of our system in \S 2,
and explaining our conventions of the system of
the Dirac equations
in \S 3, we will define the submanifold Dirac
equation following the algebraic construction in \S 4.
As the submanifold Dirac equation is on a surface in
$\EE^4$ closely related to
the generalized Weierstrass relation in \cite{Ko, M5, PP},
 we will show  Theorem 4.1 as the generalized Weierstrass
 relation with  another proof based upon \cite{M8}.
Further in \S 5, we will introduce
another submanifold Dirac operator which has
a gauge field associated with the torsion in the
sense of the Frenet-Serret relation. We will provide
Theorem 5.1
which is also
connected with the generalized Weierstrass relation.

Before finishing the Introduction,
we will comment on our theory from viewpoints
of mathematics.

Theorem 4.1 can be easily extended to
more general $k-$spin submanifold in $n$ dimensional
euclidean space $\EE^n$ as shown in \cite{M8}.
However since \cite{M8} is somewhat complicate
due to general dimensionality, in this article,
we will
restrict ourselves to four dimensional theory
in order to make theory simple
and study the effects
from the torsion in the case of
the codimention $n-k> 1$ or $(n=4, k=2)$.

However due to the potentiality, our theory
might be even complicate for differential
geometers by comparing with the theories of
\cite{Ko, PP}.
Thus we will give an answer of
a question why we persist
 the submanifold Dirac operators.

One of our motivations on this study
is to construct a
 continuous variant of
Frobenius reciprocity for linear
representation of subgroup in
a linear representation theory of
finite group \cite{S}.
This article is a first step from
\cite{M8}.
For class functions $\varphi$ and
$\psi$ over a finite group $G$ and
over its subgroup $H$ respectively, the
Frobenius reciprocity is given by,
\begin{gather}
\langle\mathrm{Res}\varphi, \psi\rangle_H
=\langle\varphi, \mathrm{Ind}\psi\rangle_G
\end{gather}
for a certain restriction map $\mathrm{Res}$,
a map of an induced representation
 $\mathrm{Ind}$ and pairings
$\langle\rangle_G$ and $\langle\rangle_H$
defined over $G$ and $H$ respectively \cite{S}.
On the other hand,
our theorem \ref{th:01} and \ref{th:02}
can be expressed by following:
For a point $\pt$ in a surface $S$ in $\EE^4$
 and for a spinor field $\psi$ over $S$ satisfying
a certain Dirac equation $\Not D \psi =0$,
we find a spinor field $\varphi$ in $S\subset \EE^4$
using data of $\EE^4$ and the relation,
\begin{gather}
\langle\varphi, \psi\rangle_S
=\langle\varphi, \psi\rangle_{\EE^4} \quad
\mathrm{at} \quad \pt,
\end{gather}
for pointwise bilinear forms at $\pt$,
$\langle\varphi, \psi\rangle_S$ and
$\langle\varphi, \psi\rangle_{\EE^4}$,
properly defined over $S$ and $\EE^4$ respectively.
Here  using a natural spin representation of
$\SO(4)$, $\langle\varphi, \psi\rangle_{\EE^4}$
gives a germ of sheaf of $T^* \EE^4$
which express the tangential component
$T^* S \subset T^* \EE^4|_S$ at the point $\pt$.
In other words,
the solution $\psi$ of the differential equation
and the pairing $\langle,\rangle_S$ at $\pt$ gives
the local data of embedding or immersion of
$S$ into $\EE^4$. These theorems give the generalized
Weierstrass relation and its essentials.
Every bilinear representation of $\SO(4)$
is uniquely identified with a spin representation
and the spinor fields can be characterized by kernel of
a Dirac operator. Thus we have searched such a Dirac
operator whose kernel has the data of immersions
for this decade \cite{M8}.
Simultaneously the fact gives answers of questions why
 the Dirac operator appears in generalized Weirestrass
relation and quaternion expresses the case of
$n=4$. It implies that our theory is natural and
 should  contrast with others.
Further we note that ${\Not D}$ is constructed by
restriction manipulations, which might
be related to Res in the group theory \cite{S}.
This author recognizers our theory as
as a continuous variant of the Frobenius reciprocity
and wishes to extend
these schemes to more general situations than \cite{M8}.
This article is its first attempt.

Further we have studied more simple
submanifold system, loops in $\EE^2$, for
this decade using the submanifold Dirac operator,
spinor representation of operator
in Frenet-Serret relation for the curvature
$k$ and the arclength $s$,
\begin{gather}
\Not D=\begin{pmatrix} \partial_s & k/2 \\
                -k/2 & \partial_s
\end{pmatrix}.
\end{gather}
This Dirac operator connects with
the various mathematics, such as
hyperelliptic function theory,
$D$-module theory,
automorphic function theory and so on
\cite{M9, M10}.
The Dirac operator plays contributes
classification of loop space of $\EE^2$
in the
category of the differential geometry,
\cite{M9, M10}. For $\Not D \psi=0$,
we rewrite it by
\begin{gather}
\begin{pmatrix} \partial_s & \\
                & \partial_s
\end{pmatrix}
\begin{pmatrix} \psi_1\\ \psi_2
\end{pmatrix}=
\begin{pmatrix}&- k/2 \\
                -k/2 &
\end{pmatrix}
\begin{pmatrix} \psi_1\\ \psi_2
\end{pmatrix},
\end{gather}
the left hand side is in the category of
differential ring whereas
 the right hand is in a category of
function space. If we restrict the right hand
side by a holomorphic function space, we
encounter the hyperelliptic functions,
via the Korteweg-de Vries or the
modified Korteweg-de Vries
equation.
The half of the curvature is connected with
the Weierstrass hyperelliptic al functions
and might be related to modular function
theory as mentioned in \cite{M9}.
The genus of hyperelliptic functions
which the Dirac operator brings us
are connected to an
infinite dimensional manifold \cite{M10}.
Further as mentioned in \cite{M1, M2},
it leads us to  index theorems.

Thus it is expected that
higher dimensional variant is
related to classification of immersions of
submanifolds in $\EE^n$
as partially mentioned in \cite{M6} for the
case of $(k,n)=(2,3)$.
From the viewpoint, it is natural
to have a question how the internal
group appears for the case of
codimension $n-k> 1$.
Thus we focus on the case of a surface
in $\EE^4$, $(n-k=2)$.

\section{Geometrical Preliminary}

In this section,
 we will give a geometrical preliminary.
Though it is not difficult to extend our theory
more general, we will concentrate our attention
on a case of a smooth surface
$S$ embedded in four dimensional euclidean space $\EE^4$,
$i: S \hookrightarrow \EE^4$.
We  identify $i(S)$ with $S$.
Since our theory is local, we construct our
theory over $S \cap U$ instead of $S$ itself
for an appropriate open set $U\subset \EE^4$
so that $S \cap U$ is topologically trivial
and its closure is compact in $\EE^4$.
For simplicity by replacing $S$ with
its piece $S \cap U$,
we assume that $S$ is homeomorphic
to $\RR^2$ and is in a compact subspace
 in $\EE^4$ hereafter.

We fix the notations of the Cartesian coordinate in
$\EE^4$ by $x:=(x^1, x^2, x^3, x^4)$.
Let $S$ be locally expressed by real parameters
$(s^1,s^2)$.
Let a tubular neighborhood of $S$ be denoted by
 $T_S$, $\pi_{T_S}:T_S \to S$.
Due  to the above assumptions, $T_S$ is homeomorphic
to $\RR^4$.
Let $q:=(q^3, q^4)$ be a normal coordinate of
 $T_S$ whose
absolute value $\sqrt{(q^3)^2+(q^4)^2}$
is the distance from the surface $S$;
$d q^\dalpha$ belongs to kernel of ${\pi_{T_S}}_*$ and
$d q^\dalpha( \partial_{s^\alpha})=0$, $(\alpha=1, 2,
\dalpha=3,4)$:
$\partial_\alpha :=\partial_{s^\alpha}
\equiv \partial/\partial s^\alpha$.
The normal bundle $T_\pt^\perp S$ is
given by $T_\pt \EE^4/T_\pt S$ at a point $\pt \in S$.
Further let
$\partial_\dalpha$
$:=\partial_{q^\dalpha}\equiv \partial/\partial q^\dalpha$,
$(\dalpha=3, 4)$.
We use the notation,
$u=(u^\mu)=(u^1,u^2,u^3,u^4)$ $:=(s^1,s^2,q^3,q^4)$.
Hereafter we will assume that
 the indices \lq\lq$\alpha$, $\beta$, $\cdots$\rq\rq\  are
for $(s^1,s^2)$, \lq\lq$\dalpha$, $\dbeta$, $\cdots$\rq\rq\  for
$(q^3,q^4)$, \lq\lq$\mu$, $\nu$, $\cdots$\rq\rq\  for $u=(s,q)$
and \lq\lq$i$, $j$, $\cdots$\rq\rq\  for the Cartesian coordinate
${x}$.

For a point of $T_S$  expressed by the
Cartesian coordinate ${x}$ in $\EE^4$
can be uniquely written by
\begin{equation*}
	{x}=\pi_{ {T_S}}{x}+q^3 {\mathbf e_3}
                                    +q^4 {\mathbf e_4},
\end{equation*}
where ${\bee_3}$ and ${\bee_4}$
are  normal unit vectors $T_{\pi_{ {T_S}}{x}}^\perp S$
 which satisfy
\begin{equation*}
    \partial_\alpha \bee_\dbeta =
    \Gamma^\beta_{\ \dbeta\alpha}
  \bee_{\beta},
\end{equation*}
for $e^i_{\ \beta}:=\partial_\beta (\pi_{T_S}{x^i})$.
A moving frame
  $E^i_{\ \mu}:=\partial_\mu x^i$, $(\mu, i=1, 2, 3, 4)$,
 in $T_S$ is expressed by
\begin{equation*}
    E^i_{\ \alpha} = e^i_{\ \alpha} +
            q^\dalpha\Gamma^\beta_{\
            \dalpha\alpha} e^i_{\ \beta},
            \quad  E^i_{\ \dalpha} = e^i_{\ \dalpha}.
\end{equation*}
In general,  more general normal unit vectors $\tilde\bee_\dalpha$
$\in T_{\pt}^\perp S$
at $\pt \in S$ obey a relation,
\begin{equation}
    \partial_\alpha \tilde \bee_\dbeta =
    \tilde \Gamma^\beta_{\ \dbeta\alpha}
  \tilde \bee_{\beta}+\tilde \Gamma^\dalpha_{\ \alpha\dbeta}
  \tilde \bee_{\dalpha}. \label{eq:frenet1}
\end{equation}
These bases can be connected with
\begin{equation}
\begin{pmatrix} \bee_3 \\ \bee_4\end{pmatrix}
=\begin{pmatrix} \cos \theta & \sin \theta \\
         -\sin \theta & \cos \theta \end{pmatrix}
\begin{pmatrix} \tilde \bee_3 \\ \tilde \bee_4\end{pmatrix},
\quad
\partial_\alpha \theta =\Gamma^3_{\ \alpha4},
\label{eq:theta}
\end{equation}
due to the relation $\Gamma^\dalpha_{\ \alpha\dbeta}
=-\Gamma^\dbeta_{\ \alpha\dalpha}$.

Thus the induced metric, $g_{T_S \mu\nu}$
$:=\delta_{i j}E^i_\mu E^j_\nu$, in $T_S$
from that in the euclidean space $\EE^4$
is given as
\begin{equation}
 g_{T_S}=\begin{pmatrix} g_{ {S_q}} & 0 \\ 0 & 1 \end{pmatrix},
  \label{eq:gTS}
\end{equation}
\begin{equation}
g_{ {S_q} \alpha\beta}
  = g_{S \alpha\beta}+
    [\Gamma_{\ \dalpha\alpha}^\Gamma g_{S\gamma\beta}+
    g_{S\alpha\gamma}\Gamma_{\ \dalpha\beta}^\gamma]q^\dalpha
    +[\Gamma_{\ \dalpha\alpha}^\delta g_{S\delta\gamma}
     \Gamma_{\ \dbeta\beta}^\gamma]q^\dalpha q^\dbeta ,
        \label{2-2}
\end{equation}
where $g_{S \alpha\beta}:=\delta_{i j}e^i_\alpha e^j_\beta$.
The determinant of the metric is expressed as
\begin{equation}
\det g_{T_S}=\rho \det g_S , \quad
\rho= (1 + \Gamma_{\dalpha \alpha}^\alpha q^\dalpha
+ \mathcal O((q^\dalpha)^2, q^3q^4) )^2 . \label{2-3}
\end{equation}

\bigskip

As we use primitive facts in sheaf theory \cite{I},
we will give our conventions.
For a fiber bundle $A$ over a paracompact
differential manifold $M$ and an
open
set $U \subset M$, let $A_M$ denote a
sheaf given by  a set of smooth local
sections of the fiber bundle $A$
and $A_M(U)\equiv\Gamma(U, A_M)$
sections of $A_M$ over $U$.
For example,
$\CC_M$ is a  sheaf given by
smooth local sections of
complex line bundle over $M$.

Further for open sets $U\subset V\subset M$,
we will denote the restriction map of a sheaf $A_M$
by $\rho_{U V}$. Using the direct limit
for $\{ U\ |\ \pt\in U \subset M\}$,
we have a {\it stalk} $A_{\pt}$ of $A_M$ by setting
$A_{\pt}\equiv \Gamma(\pt, A_M)
:=\lim_{\pt \leftarrow U}  A_M(U)$.
An element of $A_{\pt}$ is called {\it germ}.
Similarly for a compact subset $K$ in $M$,
$i_K: K \hookrightarrow M$ and
 for $\{ U\ |\ K\subset U \subset M\}$,
we have
$\Gamma(K, A_M):=\lim_{K \leftarrow U} A_M(U)$.
On the other hand, for a topological subset $N$ of
$M$, $i_N:N\hookrightarrow M$, there is an inverse sheaf,
$A_M|_N := i_N^* A_M$ given by the sections
$A_M|_N(U)=\Gamma(i_N(U), A_M)$ for $U\subset N$.
When $N$ is a compact set, {\it i.e.}, $K$,
we have an equality
$\Gamma(K, A_M)= \Gamma(K, i_K^*A_M)$
 (Theorem 2.2 in \cite{I}) and
 we identify them in this article.

We say that a sheaf $A_M$ over $M$ is
{\it soft} if and only if
for every compact subset $K \subset M$, a sheaf morphism
$\Gamma(M, A_M) \to \Gamma(K, A_M)$ is surjective.
For example, $\CC_{\RR^n}$ is soft
(Theorem 3.2 \cite{I}) and thus for a point
$\pt$ and an open set $U$, $\pt\in U \subset \RR^n$,
$\CC_{\RR^n}(U) \to \Gamma(\pt, \CC_{\RR^n})$
is surjective.
Our theory is of germs and based upon these facts.

\section{Dirac System in $\EE^4$}

For the above geometrical setting,
we will consider a Dirac equation over $T_S$
as an equation over a preHilbert space
$\mathcal H=(\Gamma_c(T_S,\Cliff_{T_S}^*)
\times \Gamma_c(T_S,\Cliff_{T_S}),
\langle,\rangle, \varphi)$.
Here  1)
$\Gamma_c(T_S, \Cliff_{T_S})$
 is a set of global sections
of the compact support Clifford module $\Cliff_{T_S}$
over $T_S$,
$\Cliff_{T_S}^*$ is a natural hermite conjugate of
$\Cliff_{T_S}$,
 2) $\langle,\rangle$ is
the L$^2$-type pairing, for $(\overline \Psi_1,\Psi_2)
\in \Gamma_c(T_S,\Cliff_{T_S}^*)\times \Gamma_c(T_S,\Cliff_{T_S})$,
\begin{equation}
	\langle\overline \Psi_1,\Psi_2\rangle = \int_{T_S} d^4 x
       \overline\Psi_1 \Psi_2,\label{eq:pairE4}
\end{equation}
and 3) $\varphi$ is the isomorphism from  $\Cliff_{T_S}$ to
$\Cliff_{T_S}^*$. Further in this article,
we will express the preHilbert space
using the triplet with the inner product
$(\circ,\cdot):=\langle\varphi \circ,\cdot\rangle$.
Here in (\ref{eq:pairE4}),
we have implicitly used another paring given by the pointwise product
for the germs
at $\pt\in T_S$, {\textit i.e.},
$ \overline\Psi_1 \Psi_2|_{\pt} \in \Gamma(\pt,\CC_{T_S})$,
which also gives us a preHilbert space
$\mathcal H^{\pt}=(\Gamma(\pt,\Cliff_{T_S}^*)\times
\Gamma(\pt,\Cliff_{T_S}), \cdot, \varphi_{\pt})$.
Here $\varphi_{\pt}$ is the  hermite conjugate operation.

Let the sheaf of the Clifford ring over $\EE^4$
and $T_S$ be denoted by $\CLIFF_{\EE^4}$ and
$\CLIFF_{T_S}$, $\CLIFF_{T_S}\equiv\CLIFF_{\EE^4}|_{T_S}$.
Since the gamma-matrix, the generator of $\CLIFF_{\EE^4}$,
depends upon the orthonormal system $\{e\}$,
we will sometimes
refer it by $\gamma_{\{e\}}(a_i e^i):=a_i\gamma_{\{e\}}(e^i)$.
In the same way, we use a representation of
the Clifford module $\Cliff_{\EE^4}$,
$\Cliff_{T_S} = \Cliff_{\EE^4}|_{T_S}$,
 using the orthonormal system $\{e\}$ as
$\Psi_{\{e\}}$.
Using the Pauli matrices,
\begin{equation*}
\tau_1:=\begin{pmatrix} 0 & 1   \\ 1    & 0  \end{pmatrix},\quad
\tau_2:=\begin{pmatrix} 0 & -\ii \\ \ii & 0  \end{pmatrix},\quad
\tau_3:=\begin{pmatrix} 1 & 0   \\ 0    & -1 \end{pmatrix},\quad
\tau_4:=\begin{pmatrix} 1 & 0   \\ 0    & 1  \end{pmatrix},
\end{equation*}
we will use the convention,
\begin{equation*}
\gamma_{\{dx\}}(dx^i):=\tau_1 \otimes \tau_i, \ (i=1,2,3),
\quad\gamma_{\{dx\}}(dx^4):=\tau_2 \otimes \tau_4.
\end{equation*}
However for abbreviation, let $\gamma^i:=\gamma_{\{d x\}}(d x^i)$.

Here the Dirac operator is given by
$\Not D_{\{d x\},x} := \gamma^i \partial_i$, and
the Dirac equation is given by
\begin{equation}
	\ii\Not D_{\{dx\},x} \Psi_{\{dx\}} = 0
         \quad \text{over}\quad T_S.
      \label{eq:D-4}
\end{equation}

Immediately we have a proposition for the solution space of
the Dirac equation.
\begin{proposition}\label{prop:01}
Let us define a set of constant section in
the Clifford module
$\Cliff_{\EE^4}$ in $\EE^4$:
\begin{equation*}
\Psi^{[1]}:=\begin{pmatrix} 1 \\0\\ 0\\ 0\end{pmatrix},
\quad
\Psi^{[2]}:=\begin{pmatrix} 0 \\ 1\\ 0\\ 0 \end{pmatrix},
\quad
\Psi^{[3]}:=\begin{pmatrix} 0 \\0\\ 1\\0 \end{pmatrix},
\quad
\Psi^{[4]}:=\begin{pmatrix} 0 \\0\\ 0\\ 1\end{pmatrix}.
\quad
\end{equation*}
 $\overline{\Psi}^{[a]}:=\varphi(\Psi^{[a]})$
is given by the hermite conjugate
of each $\Psi^{[a]}$.
\begin{enumerate}
\item They hold a relation,
\begin{equation*}
\overline \Psi^{[a]}  \Psi^{[b]} = \delta^{a,b}.
\end{equation*}
We call this relation {\textrm{orthonormal relation}}
 in this article.

\item A germ of solutions of Dirac equation (\ref{eq:D-4}) are
expressed by $\sum_a b^a(\pt) \Psi^{[a]}$ for
$b^a \in \Gamma(\pt, \CC_{\EE^4})$
at a point $\pt\in \EE^4$.
\end{enumerate}
\end{proposition}

Due to properties of the gamma-matrices,
$ \delta_{i j} \overline\Psi_1 \gamma^i\Psi_2 d x^j$
is a one-form over $T_S$.
Direct computations lead us the following Proposition
which gives the properties of Clifford module.

\begin{proposition}\label{prop:02}
Let us define a set of constant sections in Clifford
module $\Cliff_{\EE^4}$ in $\EE^4$:
\begin{equation*}
\Psi^{(1)}:=\dfrac{1}{2}\begin{pmatrix} 1 \\1\\ 1\\ 1\end{pmatrix},
\quad
\Psi^{(2)}:=\dfrac{1}{2}\begin{pmatrix} 1 \\ \ii\\ 1\\ \ii \end{pmatrix},
\end{equation*}
\begin{equation*}
\Psi^{(3)}:=\dfrac{1}{\sqrt{2}}
                    \begin{pmatrix} 1 \\0\\ 1\\0 \end{pmatrix},
\quad
\Psi^{(4)}:=\dfrac{1}{2}\begin{pmatrix} 1 \\1\\ \ii\\ \ii\end{pmatrix}.
\end{equation*}
$\overline{\Psi}^{(k)}:=\varphi(\Psi^{(k)})$
is given by the hermite conjugate
of each $\Psi^{(k)}$.
 They hold a relation,
\begin{equation*}
\delta_{ij} \overline \Psi^{(k)} \gamma^i \Psi^{(k)} d x^j
       =  dx^k \quad\mbox{(not summed over $k$)}.
\end{equation*}
We call this relation $\SO(4)${\textrm{-representation}}
 in this article.
\end{proposition}

\begin{remark}\rm{
Using a  $\CC$-valued smooth compact function
$b \in \Gamma_c(\EE^4, \CC_{\EE^4})$
over $\EE^4$ such that $b\equiv 1$ at $U \subset T_S$
and its support is in $T_S$,
$b\Psi^{[a]}$, $b\Psi^{(k)}$ and their partners
belong to $\Gamma_c(T_S,\Cliff_{T_S})$ and $\Gamma_c(T_S,\Cliff_{T_S}^*)$.
Hereafter we assume
$\Psi^{[a]}$, $\Psi^{(k)}$  and their partners
are sections of $\Gamma_c(T_S,\Cliff_{T_S})$ and
$\Gamma_c(T_S,\Cliff_{T_S}^*)$
in the sense.
}
\end{remark}

Next let us give expressions of these players
of the Dirac system in terms of the coordinate
system $u$ in $T_S$.
An orthonormal bases of $T^*T_S$ will be denoted
as $d \xi = (d \zeta^1, d \zeta^2, d q^3, d q^4)$.
Then the expressions are given by the transformations,
\begin{equation*}
   \Psi_{\{d \xi\}}(u):=\Ee^{-\Omega} \Psi_{\{dx\}}(x), \quad
   \overline\Psi_{\{d \xi\}}(u)
   := \overline\Psi_{\{dx\}}(x)\Ee^{\Omega},
\end{equation*}
\begin{equation*}
       \Ee^{-\Omega} \gamma_{\{dx\}}( d x^i )
       \Ee^{\Omega}
      ={E}_{\ \mu}^{ i} \gamma_{\{d\xi\}}( d u^\mu)
         =:{E}_{\ \mu}^{ i} \gamma^\mu.
\end{equation*}
Here $\Ee^{\Omega}$ and $\Ee^{-\Omega}$ are  sections
of the spin group sheaf $\SPIN_{T_S} \equiv\SPIN_{\EE^4}|_{T_S}$.
For $\Psi \in \Gamma_c(T_S,\Cliff_{T_S})$,
the pairing (\ref{eq:pairE4}) is expressed by
\begin{equation}
	\langle\overline{\Psi}_1,\Psi_2\rangle
	= \int (\det g_{S})^{1/2} \rho^{1/2} d^2 s d^2 q
     \  \overline\Psi_{1, \{d\zeta\}} \Psi_{2, \{d\zeta\}},
\label{eq:pairTS}
\end{equation}
and the Dirac equation (\ref{eq:D-4}) is expressed by
\begin{equation}
	\ii\Not D_{\{d\xi\},u}  \Psi_{\{d \xi\}}=0,
\quad \Not D_{\{d\xi\},u} =
\gamma^\mu(\partial_\mu +\partial_\mu \Omega).
\label{eq:DiracEqTS}
\end{equation}

Using Proposition \ref{prop:01}, we have the following
Corollary:
\begin{corollary}\label{cor:00}
For an open set $U\subset T_S$ and $\Ee^{-\Omega} \in
\Gamma(U, \SPIN_{\EE^4})$,
by letting
$\Psi_{\{d \xi\}}^{[a]}:=\Ee^{-\Omega} \Psi^{[a]}
\in \Gamma(U, \Cliff_{T_S})$
and $\overline\Psi_{\{d \xi\}}^{[a]}
   := \overline\Psi^{[a]}\Ee^{\Omega}
\in \Gamma(U, \Cliff_{T_S}^*)$ over U,
the  orthonormal relation holds
\begin{equation}
\overline{\Psi}_{\{d \xi\}}^{[a]}\Psi_{\{d \xi\}}^{[b]}
= \delta^{a,b}
\quad \mbox{at}\quad U\subset T_S.
\label{eq:orthTS}
\end{equation}

Inversely for given such orthonormal bases
$\Psi_{\{d \xi\}}^{[b]}\in \Gamma(U, \Cliff_{T_S})$ and
$\varphi(\Psi_{\{d \xi\}}^{[b]})$
satisfying (\ref{eq:orthTS}),
the relation,
$$
\Psi_{\{d \xi\}}^{[a]}=\Ee^{-\Omega} \Psi^{[a]}
\quad \mbox{for}\quad a=1,2,3,4,
$$
completely characterizes the spin matrix $\Ee^{-\Omega}
\in \Gamma(U, \SPIN_{T_S})$.
\end{corollary}

Proposition \ref{prop:02} gives the following
Corollary:
\begin{corollary}\label{cor:01}
For an open set $U\subset T_S$ and $\Ee^{-\Omega} \in
\Gamma(U, \SPIN_{\EE^4})$,
by letting
$\Psi_{\{d \xi\}}^{(i)}:=\Ee^{-\Omega} \Psi^{(i)}
\in \Gamma(U, \Cliff_{T_S})$
and $\overline\Psi_{\{d \xi\}}^{(i)}
   := \overline\Psi^{(i)}\Ee^{\Omega}
\in \Gamma(U, \Cliff_{T_S}^*)$ over $U$,
the  $\SO(4)$-representation holds
$$
g_{T_S \mu\nu}\overline \Psi_{\{d \xi\}}^{(i)}
      \gamma^\mu\Psi_{\{d \xi\}}^{(i)} d u^\nu = d x^i
\quad \mbox{at}\quad U\subset T_S
\quad \mbox{(not summed over i)}.
$$
\end{corollary}

\section{ Submanifold Dirac Operator over $S$ in $\EE^4$}

In this section, we will define the submanifold Dirac
operator over  $S$ in $\EE^4$ and investigate
its properties.
In the  papers in \cite{M1, M2, M3, M4, M5, MT},
we add a mass type potential in (\ref{eq:DiracEqTS}),
 which confines a particle in the
tubular neighborhood $T_S$;
the mass potential makes the support of the
Clifford module
in $T_S$. After taking a squeezing limit
of the mass potential, we decompose the normal and the
tangential modes, suppress the normal mode,
 and obtain the submanifold
Dirac equation as an effective equation for low energy
states.
Instead of the scheme, we will choose another construction
and give a novel
definition of the Dirac operator as in Definition
\ref{SubD} \cite{M8}.

Let us consider such a Dirac particle algebraically.
Confinement of the particle into a surface requires
that the momentum and position of the particle vanish.
In order to realize  the vanishing momentum, we
wish to consider kernel of $\partial_\dalpha$.
However $p_\dalpha:=\sqrt{-1}\partial_\dalpha$ is not
self-adjoint in general
due to the existences $\rho$ in (\ref{eq:pairTS}).

For an operator $P$ over $\Cliff_{T_S}$, let $Ad(P)$ be defined by
the relation,  $\langle\overline\Psi_1,P \Psi_2\rangle_
=\langle\overline\Psi_1Ad(P), \Psi_2\rangle$ if exists.
Further for $\Psi\in\Gamma_c(T_S,\Cliff_{T_S})$, $P^*$ is defined by
$P^*\Psi = \varphi^{-1}(\varphi(\Psi)Ad(P))$.
If $p_\dalpha$ is not self-adjoint, the kernel
of $p_\dalpha$
is not isomorphic to the kernel of $Ad(p_\dalpha)$.
Thus the kernel of $p_\dalpha$ cannot become
a preHilbert space and $\varphi_{\pt}$ or $\varphi_{\pt}^{-1}$
is not well-defined there.
It means that $\SO(4)$-representation,
Corollary \ref{cor:01}, which should be regarded as
a fundamental properties of the
Clifford module, will neither be
well-defined.

Accordingly we introduce another preHilbert space
$\mathcal H'
\equiv ( \Gamma_c(T_S,\overwave\Cliff_{T_S}^*)
\times  \Gamma_c(T_S,\overwave\Cliff_{T_S}),
\langle,\rangle_\sa, \tilde\varphi)$ so that
$p_\dalpha$'s become  self-adjoint operators
there.
Using the half-density
(Theorem 18.1.34 in \cite{H}),
we construct
{\textit self-adjointization}:
$\eta_\sa : \mathcal H \to \mathcal H'$
by,
\begin{equation*}
	\eta_\sa(\overline \Psi) := \rho^{1/4} \overline\Psi, \quad
	\eta_\sa(\Psi) := \rho^{1/4} \Psi, \quad
 	\eta_\sa(P) :=\rho^{1/4} P \rho^{-1/4}.
\end{equation*}
Here since $\rho$ does not vanish in $T_S$, $\eta_\sa$ gives
an isomorphism
$\eta_\sa:
\Cliff_{T_S}^* \times \Cliff_{T_S}
\to \overwave \Cliff_{T_S}^* \times\overwave \Cliff_{T_S}$.
For $(\overline\Psi_1,\Psi_2)\in  \Gamma_c(T_S,\tilde\Cliff_{T_S}^*)
 \times  \Gamma_c(T_S,\tilde\Cliff_{T_S})$,
by letting $\tilde \varphi:= \eta_\sa \varphi \eta_\sa^{-1}$,
the pairing is defined by
\begin{equation}
	\langle\overline\Psi_1,\Psi_2\rangle_\sa := \int_{T_S}
 (\det g_{S})^{1/2} d^2 s d^2 q\  \overline\Psi_1 \Psi_2.
\label{eq:pairS}
\end{equation}
Here we have the properties of $\eta_{\mathrm{sa}}$
that 1)
$\langle\circ,\cdot\rangle_\sa
=\langle\eta_\sa^{-1}\circ,\eta_\sa^{-1}\cdot\rangle$,
 2) for an operator $P$ of $\Cliff_{T_S}$,
$\eta_\sa(P)=\eta_\sa P \eta_\sa^{-1}$,
and 3)
 $p_\dalpha$'s themselves become self-adjoint
in $\mathcal H'$, {\textit i.e.},
$p_\dalpha =p_\dalpha^*$.
The self-adjointization is not a unitary operation
in some sense
because due to the operation, the inner product changes from
$\langle\varphi\circ,\cdot\rangle$
to $\langle\tilde\varphi\circ,\cdot\rangle_\sa$
if we regard them as inner products for
 $\Gamma_c(T_S,\tilde\Cliff_{T_S})
 \times  \Gamma_c(T_S,\tilde\Cliff_{T_S})$.
Due to the trick,
$p_\dalpha$'s become self-adjoint.

Noting $\rho=1$ at a point in $S$,
Corollaries \ref{cor:00} and  \ref{cor:01} lead the following
lemma.
\begin{lemma}
\begin{enumerate}
\item
For $(\overline\Psi, \Psi) \in \Gamma(S, \Cliff_{T_S}^*) \times
\Gamma(S, \Cliff_{T_S})$,
$\eta_\sa(\Psi)=\Psi$ and
$\eta_\sa(\overline\Psi) = \overline\Psi$ at $S$.

\item
For the quantities defined in Corollary \ref{cor:00},
the orthonormal relation holds:
\begin{equation*}
\eta_\sa(\overline{\Psi}_{\{d \xi\}}^{[a]})
\eta_\sa(\Psi_{\{d \xi\}}^{[b]})
=\delta^{a,b} \quad \mbox{at}\quad S.
\end{equation*}

\item
For the quantities defined in Corollary \ref{cor:01},
by letting
$\Phi^{(i)}:=\eta_\sa(\Psi_{\{d \xi\}}^{(i)})$
and
$\overline{\Phi}^{(i)}:=
\eta_\sa(\overline{\Psi}_{\{d \xi\}}^{(i)})$,
the  $\SO(4)$-representation  holds
\begin{equation*}
g_{T_S \mu\nu}\overline{\Phi}^{(i)}
      \gamma^\mu\Phi^{(i)} d u^\nu = d x^i
\quad \mbox{at}\quad S\quad \mbox{(not summed over $i$)}.
\end{equation*}
\end{enumerate}
\end{lemma}

We have the following proposition.

\begin{proposition}
By letting $p_q:=a_3 p_3 + a_4 p_4$ for real
generic numbers $a_3$
and $a_4$,
the projection,
\begin{equation*}
	\pi_{p_q}: \overwave \Cliff_{T_S}^* \times
             \overwave \Cliff_{T_S} \to
          \Ker( Ad(p_q))\times \Ker (p_q),
\end{equation*}
induces the projection in the preHilbert space \cite{A},
 {\textit i.e.},
\begin{enumerate}
\item
$\tilde\varphi|_{\Ker(p_q)}: \Ker (p_q) \to
\Ker (Ad(p_q))$ is isomorphic.
We simply express $\tilde\varphi|_{\Ker(p_q)}$
by $\tilde \varphi$ hereafter.

\item $\mathcal H_{p_q}:=(
\Gamma_c(T_S, \Ker (Ad(p_q)))\times
\Gamma_c(T_S, \Ker (p_q)),
\langle,\rangle_\sa,\tilde\varphi)$ is a preHilbert space.

\item $\varpi_{p_q}:=\pi_{p_q}|_{\overwave \Cliff_{T_S}}$,
$\varpi_{p_q}=\varpi_{p_q}^2=\varpi_{p_q}^*$ in
$\mathcal H_{p_q}$.

\item $\varpi_{p_q}$ induces a natural restriction of pointwise
multiplication for a point in $T_s$,
$\mathcal H^{\pt}_{p_q}:=(
\Gamma(\pt, \Ker (Ad(p_q)))\times
\Gamma(\pt, \Ker (p_q)), \cdot,
\tilde\varphi_{\pt})$ becomes a preHilbert space.
The hermite conjugate map $\tilde\varphi_{\pt}$ is
still an isomorphism.
\end{enumerate}
\end{proposition}

\begin{proof}
Since $p_\dalpha$ is self-adjoint, $\Ker (p_q)=\Ker (p_q^*)$
and $\Ker(p_q)$ is isomorphic to
$ \Ker (Ad(p_q))$, {\textit i.e.},
$\varphi(\varpi_{p_q}\Psi) = \varphi(\Psi)Ad(\varpi_{p_q})$.
$\varpi_{p_q}^*\Psi = \varphi^{-1}(\varphi(\Psi)Ad(\varpi_{p_q}))$
gives $\varpi_{p_q}=\varpi_{p_q}^*$.
\end{proof}

\bigskip

Since $T_S$ is homeomorphic to $\RR^4$,
$\CC_{T_S}$  is soft (Theorem 3.1 in \cite{I}).
Hence we have the following proposition.
\begin{proposition}\label{prop:41}
$\Cliff_{T_S}$ is soft.
\end{proposition}

\begin{proof}
Due to Proposition \ref{prop:01} (2), $\Cliff_{T_S}$ is
 a sheaf of $\CC_{T_S}$ vector bundle.
 From the proof of Theorem 3.2 in \cite{I},
 it is justified.
\end{proof}

Due to the Proposition \ref{prop:41},
for a point $\pt$ in $S$, $\pt\in U\subset T_S$
and for
a germ $\Psi_{\pt}\in\Gamma(\pt, \Cliff_{T_S})$,
there exists $\Psi_c\in\Gamma_c(T_S, \Cliff_{T_S})$
$\Psi_o\in\Gamma(U, \Cliff_{\EE^4})$ such that
\begin{gather*}
      \Psi_{\pt} = \Psi_c,
\quad
      \Psi_{\pt} =\Psi_o,\quad \mbox{at} \quad \pt.
\end{gather*}
Thus when we deal with an element of
$\Gamma(\pt, \Cliff_{T_S})$, we need not distinguish
which it comes from $\Gamma_c(T_S, \Cliff_{T_S})$
or $\Gamma(U, \Cliff_{\EE^4})$.

\begin{remark}\label{re:4-1}\rm{
At a point $\pt$ in S, we can find
$(\overline{\Phi}^{(i)},\Phi^{(i)})$ in
$\mathcal H^{\pt}_{p_q}$ obeying the
 $\SO(4)$-representation,
\begin{equation*}
g_{T_S \mu\nu}\overline{\Phi}^{(i)}
      \gamma^\mu\Phi^{(i)} d u^\nu = d x^i
\quad \mbox{at}\quad \pt\quad
\mbox{(not summed over $i$)}.
\end{equation*}
because 1) we can easily find such an element in $\mathcal H'$
and 2)  extend its domain to vicinity of $S$
so that its value preserves for the normal direction, {\textit, i.e.},
$\partial_\dalpha \Phi^{(i)}=0$ and
$\overline{\Phi}^{(i)}Ad(\partial_\dalpha)=0$;
$(\overline{\Phi}^{(i)},\Phi^{(i)})$ belongs to
$\Ker(p_q)^*\times \Ker(p_q)$.
}
\end{remark}

Since we kill a normal translation freedom in $\mathcal H_{p_q}$,
we can choose a position $q$ and make $q$ vanish.
Thus we will give our definition of the submanifold
Dirac operator.

\begin{definition}\label{SubD}{\textrm
We define the submanifold Dirac operator for the
surface $S$ in $\EE^4$ by,
\begin{equation*}
	\Not D_{S \hookrightarrow \EE^4}
         :=\eta_\sa(\Not D)|_{\Ker( p_q)}|_{q=0},
\end{equation*}
as an endomorphism of
Clifford submodule $\Ker(p_q)|_S$ $\subset$
$\overwave\Cliff_{T_S}|_S$, {\it i.e.},
$
{{\Not D_{S \hookrightarrow \EE^4}}}:
\Ker(p_q)|_S \to  \Ker(p_q)|_S
$.
}
\end{definition}

Here we note that the first restriction
$|_{\Ker( p_q)}$ is as
an operator but the second one
$|_{q=0}$ is given by $\rho_{S, T_S}$,
which is given by a direct limit
$\rho_{S, T_S} :=\lim_{S \rightarrow U} \rho_{U, T_S}$.

\vskip 1.0  cm

Here we will connect the $\CLIFF_{T_S}|_S$
with the sheaf of proper Clifford ring $\CLIFF_S$  over $S$.
We will fix the orthonormal base $\{d \zeta\}
\equiv (d \zeta^\alpha)$ associated with the local coordinate
$(d s^\alpha)$ and the gamma-matrix as a generator
in $\CLIFF_S$ by
$\gamma_{S, \{d \zeta\}}( d \zeta^\alpha)$.
For abbreviation,
$\tilde\sigma^\alpha :=\gamma_{S, \{d \zeta\}}(d \zeta^\alpha)$
and
$\sigma^\alpha :=\gamma_{S, \{d \zeta\}}(d s^\alpha)$.

We have an inclusion as vector space for generators,
\begin{equation*}
\iota_g:\CLIFF_{S} \ni \tilde\sigma^\alpha
\mapsto \tau_1 \otimes
\tilde\sigma^\alpha \in
            \CLIFF_{T_S}|_S.
\end{equation*}
Let $\iota_g (\tilde\sigma^\alpha\tilde\sigma^\beta):=
\iota_g (\tilde\sigma^\alpha )
\iota_g (\tilde\sigma^\beta )$,
$\iota_g (\tilde\sigma^\alpha\tilde\sigma^\beta\tilde\sigma^\gamma):=
\iota_g (\tilde\sigma^\alpha )
\iota_g (\tilde\sigma^\beta )\iota_g (\tilde\sigma^\gamma)$ and so on.
This does not become the homomorphism
between the Clifford rings whereas the natural
ring homeomorphism is given by
\begin{equation*}
\iota_r: \CLIFF_{S}\ni c \mapsto 1 \otimes c \in
 \CLIFF_{T_S}|_S.
\end{equation*}
However the inclusion $\iota_g$
generates the homomorphism of the
spin groups because
$
\iota_g (\tilde\sigma^\alpha \tilde\sigma^\beta )
= \iota_r (\tilde\sigma^\alpha \tilde\sigma^\beta ).
$
A spin matrix $\exp(\Omega_S) \in \Gamma(\pt,\SPIN_S) $,
of the spin group sheaf $\SPIN_S$ properly
defined over $S$,
 is given by
$
\exp(\Omega_S)=\exp(a_{\alpha \beta }
\tilde\sigma^\alpha \tilde\sigma^\beta )$.
On the other hand, a germ $\exp(\Omega)$ of $\SPIN_{T_S}$ at
a point $\pt$ in S is given by
 $
\exp(\Omega)=\exp(a_{\mu \nu }
\gamma^\mu \gamma^\nu )
$. Thus $\iota_g$ and $\iota_r$ induce the
natural inclusion of $\SPIN_S$ into
$\SPIN_{T_S}$ as a sheaf morphism by
 $
\exp(\Omega)=\exp(a_{\alpha \beta }
(1\otimes \tilde\sigma^\alpha \tilde\sigma^\beta) )$.

Using these facts, we will give explicit form
of the submanifold Dirac operator, which was obtained in
\cite{M5} using a mass potential.
\begin{proposition}
The submanifold Dirac operator of the surface $S$ in
$\EE^4$ can be expressed by
\begin{equation}
	\Not D_{S \hookrightarrow \EE^4}
         =
	\iota_g(\sigma^\alpha \nabla_\alpha)
 +  \frac{1}{2}\gamma^3\Gamma^\alpha_{\ 3\alpha}
+  \frac{1}{2}\gamma^4\Gamma^\alpha_{\ 4\alpha},
\label{eq:DiracSE4}
\end{equation}
where $\nabla_\alpha$ is the proper spin connection over
$S$ and $\gamma^\dalpha :=\gamma_{\{d\xi\}}(d q^\dalpha)$.
\end{proposition}

\begin{proof}
First we note that $\eta_\sa(\Not D_{\{d\xi\},u})$ has
a decomposition,
\begin{equation*}
\eta_\sa(\Not D_{\{d\xi\},u})=
\Not{\mathbb D}_{\{d\xi\},u}^\parallel+
                    \Not {\mathbb D}_{\{d\xi\},u}^\perp,
\end{equation*}
where $\Not {\mathbb D}_{\{d\xi\},u}^\perp:=\gamma^\dalpha \partial_\dalpha$
and $\Not {\mathbb D}_{\{d\xi\},u}^\parallel$ does not include the normal
derivative $p_\dalpha$. $\Not D_{\{d\xi\},u}^\perp$ vanishes
at $\Ker(p_q)$. Due to the constructions,
$\iota_g(\sigma^\alpha)$ and $\gamma^\dalpha$ become generator
of the $\CLIFF_{T_S}$ at sufficiently vicinity of $S$.
The geometrically independency due to (\ref{eq:gTS}) and
direct computations give above the result.
\end{proof}

Now we will give the first theorem:
\begin{theorem}\label{th:01}
Let  a point $\pt$ in $S$ be
 expressed by the Cartesian coordinate $(x^i)$ and
$\CC^4_{S}$ sheaf of complex vector bundle over
$S$ with rank four.
A set of germs of $\Gamma(\pt, {\CC^4_{S}})$
satisfying the
submanifold Dirac equation,
\begin{equation*}
	\ii\Not D_{S \hookrightarrow \EE^4} \psi
= 0 \quad \text{at}\quad \pt,
\end{equation*}
is given by $\{b_a\psi^{[a]} \ |\ a=1, 2, 3, 4, \ b_a \in \CC \}$
 such that
$$
	\varphi_{\pt}(\psi^{[a]})\psi^{[b]}= \delta_{a,b}
\quad \text{at}\quad \pt.
$$
At the point $\pt$, there exists a spin matrix
$\Ee^{\Omega}\in\Gamma(\pt,\SPIN_{\EE^4})$ satisfying
$\psi^{[a]}=\Ee^{-\Omega}\Psi^{[a]}$ $(a=1,2,3,4)$.
We define
$\psi^{(i)}:=\Ee^{-\Omega}\Psi^{(i)}$ and
$\overline  \psi^{(i)}:=\overline \Psi^{(i)}\Ee^{-\Omega}$
$(i=1,2,3,4)$ at the point.
Then the following relation holds:
\begin{equation}
 g_{S, \alpha,\beta}\overline\psi^{(i)} [\iota_g
(\sigma^\beta))] \psi^{(i)}
           =\partial_{s^\alpha} x^i,
\quad \text{at}\quad \pt,
   \quad\mbox{(not summed over $i$)}.
\end{equation}
\end{theorem}

\begin{remark}\label{re:4-2}
\rm{
\begin{enumerate}
\item As our theory is a local theory, this theorem can be
extended to any surfaces immersed in $\EE^4$.

\item
If the surface is conformal,
this theorem means the generalized
Weierstrass relation in
$\EE^4$ given by Konopelchenko \cite{Ko}
and Pedit and Pinkall \cite{PP}
as mentioned in \cite{M5}.

\item
This can be easily generalized to a $k$ submanifold $S^k$
immersed in the $n$-euclidean space $\EE^n$. In the
above statements, the index \lq\lq$a=1,\cdots,4$\rq\rq
should be replaced to
\lq\lq$a=1,2,\cdots,2^{[n/2]}$\rq\rq,
\lq\lq$i$\rq\rq to \lq\lq$i=1,\cdots,n$\rq\rq,
\lq\lq$\alpha=1,\cdots,k$\rq\rq, and
\lq\lq$\dalpha=k+1,\cdots, n$\rq\rq \cite{M8}.
\end{enumerate}
}
\end{remark}

\begin{proof}
Since $\Not D_{S \hookrightarrow \EE^4}$ is the
four rank first order differential operator and
has no singularity over $S$ due to the construction,
a germ of its kernel in $\Gamma(\pt, \CC_{\EE^4}^4)$
 is given by four dimensional
vector space at each point of $S$.
Since $\Not D_{S \hookrightarrow \EE^4}$
is defined as an endomorphism of $\Ker(p_q)$
and  $\Ker(p_q)$ contains the zero section, the germ of
kernel of the Dirac operator,
$\Ker($ ${\Not D}_{S \hookrightarrow \EE^4})$,
 is a submodule
of $\Ker(p_q)|_S$.
Let $\Not{\mathbb D}^\perp := \gamma^\dalpha \partial_\dalpha$
at $S$.
From the construction, we have
$$
	\Not D_{S \hookrightarrow \EE^4} + \Not{\mathbb D}^\perp
       = \eta_\sa(\Not D_{\{d\xi\},u}) |_S.
$$
Hence the solution of
$\Not D_{S \hookrightarrow \EE^4}$ becomes a solution of
$\eta_\sa(\Not D_{\{d\xi\},u}) |_S$.
Noting Remark \ref{re:4-1}, $\tilde \varphi_{\pt}$ is an isomorphism
and $\mathcal H^{\pt}_{p_q}$ gives
$\SO(4)$-representation. Thus we prove it.
\end{proof}

\section{ Submanifold Dirac Operator over $S$ in $\EE^4$
with Torsion}

For the transformations (\ref{eq:theta}), we have
the relation,
\begin{equation*}
\begin{pmatrix} \Gamma^\beta_{\ 3\alpha} \\
                \Gamma^\beta_{\ 4\alpha}\end{pmatrix}
=\begin{pmatrix} \cos \theta & \sin \theta \\
         -\sin \theta & \cos \theta \end{pmatrix}
\begin{pmatrix} \tilde\Gamma^\beta_{\ 3\alpha} \\
                \tilde\Gamma^\beta_{\ 4\alpha}\end{pmatrix}.
\label{eq:theta2}
\end{equation*}
Now let us choose $\vartheta$ in $\{\theta \in [0,2\pi]\}$
as
\begin{equation*}
\Gamma^\alpha_{\ 3\alpha}\sin\vartheta +
\Gamma^\alpha_{\ 4\alpha}\cos\vartheta =0,
\end{equation*}
and define
\begin{equation*}
\hat\Gamma^\alpha_{\ 3\alpha}:=
\Gamma^\alpha_{\ 3\alpha}\cos\vartheta +
\Gamma^\alpha_{\ 4\alpha}\sin\vartheta,
\quad
\hat\Gamma^\alpha_{\ 4\alpha}=0,
\quad\hat\Gamma^3_{\ \alpha4}:=\partial_\alpha\vartheta.
\end{equation*}
We call $\hat\Gamma^3_{\ \alpha4}$
{\textit{torsion}} in this system
in the sense of the Frenet-Serret relation.
It obeys the relation,
\begin{equation*}
	\Gamma^\alpha_{\ 3\alpha}
         = \hat\Gamma^\alpha_{\ 3\alpha} \cos \vartheta, \quad
        \Gamma^\alpha_{\ 4\alpha}
        = -\hat\Gamma^\alpha_{\ 3\alpha} \sin \vartheta.
\end{equation*}
The Dirac operator (\ref{eq:DiracSE4}) can be
expressed by
\begin{equation*}
	\Not D_{S \hookrightarrow \EE^4}
         =
	\gamma^\alpha \nabla_\alpha +
             \frac{1}{2} \gamma^3 \hat\Gamma^\alpha_{\ 3\alpha}
                 \Ee^{\sigma^{34} \vartheta},
\end{equation*}
where $\sigma^{34}:=\gamma^3\gamma^4$.
This type operator appears in \cite{M2}.
For the gauge transformation,
\begin{equation*}
	\Not D_{S \hookrightarrow \EE^4}^{\vartheta}
        := \Ee^{-\sigma^{34} \vartheta}
            \Not D_{S \hookrightarrow \EE^4}
           \Ee^{\sigma^{34} \vartheta},
\end{equation*}
we have
\begin{equation*}
	\Not D_{S \hookrightarrow \EE^4}^{\vartheta}
         =	\gamma^\alpha( \nabla_\alpha
                     +\gamma_\alpha \sigma^{34}
     \hat\Gamma^3_{\ \alpha4} )
             +  \frac{1}{2}\gamma^3  \hat\Gamma^\alpha_{\ 3\alpha}.
\end{equation*}
Here $\gamma_\alpha := g_{S \alpha,\beta}\gamma^\beta$.
We call this operator
{\textit{gauged submanifold Dirac operator}}.
This is a generalization of the Takagi-Tanzawa Schr{\"o}dinger operator
in \cite{TT}.

Let us show our main theorem in this article:
\begin{theorem}\label{th:02}
Fix  a point $\pt$ in $S$
 expressed by the Cartesian coordinate $(x^i)$.
The germs in $\Gamma(\pt,\CC^4_{S})$ satisfying the
gauged submanifold Dirac equation,
\begin{equation*}
	\ii\Not D_{S \hookrightarrow \EE^4}^{\vartheta} \psi
= 0 \quad \text{at}\quad \pt,
\end{equation*}
is given by $\{b_a\psi^{[a]} \ |\ a=1, 2, 3, 4, \ b_a\in \CC \}$
such that
$$
	\varphi_{\pt}(\psi^{[a]})\psi^{[b]}= \delta_{ab}.
$$
There exists a spin matrix
$\Ee^{\hat\Omega}\in \Gamma(\pt, \SPIN_{\EE^4})$ satisfying
$\psi^{[a]}=\Ee^{-\hat\Omega}\Psi^{[a]}, (a=1,2,3,4)$.
By defining
$\psi^{(i)}:=\Ee^{-\hat\Omega}\Psi^{(i)}$ and
$\overline  \psi^{(i)}:=\overline \Psi^{(i)}\Ee^{-\hat\Omega}$,
$(i=1,2,3,4)$, the following relation holds:
\begin{equation}
 g_{S, \alpha,\beta}\overline\psi^{(i)} [\iota_g
(\sigma^\beta))] \psi^{(i)}
           =\partial_{s^\alpha} x^i
   \quad\mbox{(not summed over $i$)}.
\end{equation}
\end{theorem}

\begin{proof}
Let $\Ee^{-\hat\Omega}=\Ee^{-\Omega}\Ee^{-\sigma^{34}}$.
Due to the proof of theorem \ref{th:01}, we can prove it.
\end{proof}

\begin{remark}
\rm{
\begin{enumerate}
\item When we extend the gauged submanifold Dirac operator
to that over a compact surface, there appears a problem
whether $\vartheta$ can be globally defined or not.
$\hat\Gamma^3_{\ \alpha4}$ appears an associated
 gauge field.

\item
If $\hat\Gamma^\alpha_{\ 3\alpha}$ is constant case,
it can be regarded as a mass of the Dirac particle
and further the torsion plays a role of the
$U(1)$-gauge field. This is very interesting
from physical viewpoint. Even though in the string theory,
the extra dimensions are connected with gauge fields,
it is surprising that the gauge field appears as
the torsion of the
submanifold.

\item
It is not difficult to extend our theory to
that in $k$-submanifold in $\EE^n$. Then it is expected
that there appears a $\SO(n-k)$-gauge field.

\item The determinants of the
submanifold Dirac operators are related to the geometrical
properties as in \cite{M1, M2, M4}. It should be expected
that the gauged submanifold Dirac operator also brings
us to the data of submanifold such as index theorem.

\item The Dirac operator of a conformal surface in $\EE^3$
are related to the extrinsic string as in \cite{M4, M6}.
The gauged submanifold Dirac operator might also be connected with
the extrinsic string.
\end{enumerate}
}
\end{remark}

\vskip 0.5 cm


\begin{thebibliography}{9}

\bibitem{A}
    H. Araki,
    \book{Mathematical Theory of Quantum Fields},
(International Series of Monographs on Physics, No. 101)
  Oxford, 
 1999.


\bibitem{Bj}
 J-E. Bj{\"o}rk,
    \book{Analytic $\mathcal D$-Modules and Applications},
     Kluwer, 
   1992.




\bibitem{BJ}
M.~Burgess and B.~Jensen,
\paper{Fermions near two-dimensional surfaces},
\jour{Phys. Rev. A}
\vol{48}  \yr{1993}, \pages 1861-1866.


\bibitem{dC}
R. C. T.  da Costa,
\paper{Quantum  mechanics of a constrained particle},
\jour{Phys. Rev A} \vol{23} \yr{1981}, \pages 1982-1987.


\bibitem{H}
L. H{\"o}rmander,
\book{The analysis of linear partial differential
operators III},
Springer-Verlag,
 1985.

\bibitem{I}
 B. Iversen,
    \book{Cohomology of Sheaves},
    Springer-Verlag, 
  1986.

\bibitem{INTT}
M. Ikegami, Y. Nagaoka, S. Takagi and T. Tanzawa,
\paper{Quantum Mechanics of a Particle on a Curved Surface -
Comparison of Three Different Approaches-},
\jour{Prog. Theor. Phys.} \vol{88} \yr{1992}, \pages 229-249.


\bibitem{JK}
  H.~Jensen and H.~Koppe,
\paper{Quantum Mechanics with Constraints},
\jour{Ann. Phys.} \vol{63} \yr{1971}, \pages{586-591}.

\bibitem{K}
 D. Krej{\u c}i{\u r}\'ik,
\paper{Quantum strips on surfaces},
 \jour{J. Geom. Phys.} \vol{45}
\yr{2003}, \pages{203-217}.


\bibitem{Ko}
B. G. Konopelchenko,
\paper{Weierstrass representations for surfaces in 4D spaces
and their integrable deformations via DS hierarchy},
\jour{Ann. Global Analysis and Geom.}
\vol{18} \yr{2000}, \pages 61-74.


\bibitem{M1}
S.~Matsutani,
\paper{The Relation between the Modified Korteweg-de
Vries Equation  and Anomaly of Dirac Field on a Thin Elastic Rod},
\jour{  Prog.Theor.  Phys.}
       \vol{5} \yr{1994}, \pages  1005-1037.

\bibitem{M2}
\bysame,
\paper{Anomaly on a Submanifold System:
 New Index Theorem related to a Submanifold System},
 \jour{J.  Phys.  A: Gen. \& Math.}  \vol{28}
\yr{1995}, \pages{1399-1412}.


\bibitem{M3}
 \bysame,
       \paper{Constant Mean Curvature Surface and Dirac Operator},
       \jour{J. Phys.  A: Gen. \& Math.} \vol{30}
        \yr{1997}, \pages  4019-4029.


\bibitem{M4}
\bysame,
      \paper{Immersion Anomaly of Dirac Operator on Surface in ${\mathbb R}^3$}
      \jour{Rev. Math. Phys.}  \vol{2}
      \yr{1999}, \pages 171-186.

\bibitem{M5}
\bysame,
      \paper{Dirac Operator of a Conformal Surface Immersed in ${\mathbb R}^4$:
      Further Generalized Weierstrass Relation}
      \jour{Rev. Math. Phys.} \vol{12} \yr{2000}, \pages {431-444}.

\bibitem{M6}
 \bysame,
      \paper{On Density of State of Quantized Willmore Surface
       :A Way to a Quantized Extrinsic String in ${\mathbf R}^3$},
       \jour{J. Phys. A: Math. Gen.}  \vol{31} \yr{1998}, \pages
       3595-3606.


\bibitem{M9}
\bysame,
\paper{Hyperelliptic Loop Solitons with Genus $g$:
          Investigations of a Quantized Elastica}
        \jour{J. Geom. Phys.}  \vol{43} \yr{2002} 146-162.

\bibitem{M10}
 \bysame,
         \paper{On the Moduli of a Quantized Elastica in
       $\mathbb{P}$
and KdV Flows: Study of Hyperelliptic Curves as an
Extension of Euler's Perspective of Elastica I},
         \jour{Rev. Math. Phys.} \vol{15} \yr{2003}, No.5.

\bibitem{M7}
\bysame,
\paper{On an Essential of Submanifold Quantum Mechanics},
\jour{J. Geom. Symm. Phys.} \vol{1} \yr{2003}.


\bibitem{M8}
 \bysame,
\paper{Generalized Weierstrass Relation for a Submanifold $S^k$ in
${\mathbf E}^n$
 Coming from Submanifold Dirac Operator},
\jour{ Adv. Stud. Pure Math.} (2004).


\bibitem{MT}
 S.~Matsutani and H.~Tsuru,
\paper{Physical relation between
quantum mechanics and soliton on a thin elastic rod},
\jour{Phys.  Rev.  A}
\vol{46} \yr{1992}, \pages 1144-1147.

\bibitem{PP}
  F. Pedit and U. Pinkall,
\paper{Quaternionic Analysis on Riemann Surfaces and Differential Geometry},
\jour{Doc. Math. J. DMV} Extra Vol. ICM II
(1999), 389-400.

\bibitem{S}
J-P. Serre,
\book{Linear Representations of Finite Group},
Springer,
1977.

\bibitem{Ta}
 I. A. Taimanov,
\paper{The Weierstrass representation of closed surfaces in $ R\sp 3$},
\jour{Funct. Anal. Appl.} \vol{32} (1998), 258--267.

\bibitem{TT}
 S. Takagi and T. Tanzawa,
\paper{Quantum mechanics of a particle confined to a twisted ring},
\jour{Prog. Theor. Phys.} \vol{87} \yr{1992},
\pages 561-568.



\end{thebibliography}

\bigskip

\leftline{{Shigeki Matsutani}}

\leftline{{e-mail:rxb01142@nifty.com}}

\leftline{{8-21-1 Higashi-Linkan,}}

\leftline{{Sagamihara 228-0811}}

\leftline{{JAPAN}}

\end{document}